\documentclass{article}
\usepackage[T1]{fontenc} 
\usepackage[utf8]{inputenc} 
\usepackage{ismir,amsmath,cite,url}
\usepackage{graphicx}
\usepackage{color}
\usepackage{booktabs}
\usepackage{enumitem}
\usepackage{subfigure}
\usepackage[bookmarks=false,hidelinks]{hyperref}

\usepackage{lineno}

\title{Emotion-driven Piano Music Generation via Two-stage Disentanglement and Functional Representation}

\multauthor
{Jingyue Huang$^1$ \hspace{1cm} Ke Chen$^1$ \hspace{1cm} Yi-Hsuan Yang$^2$} { 
$^1$Department of Computer Science and Engineering, UC San Diego \\
$^2$Department of Electrical Engineering, National Taiwan University \\
{\tt\small \{jih150, knutchen\}@ucsd.edu, yhyangtw@ntu.edu.tw}
}

\def\authorname{J. Huang, K. Chen, and Y. Yang}

\begin{document}
\maketitle
\begin{abstract}

Managing the emotional aspect remains a challenge in automatic music generation.
Prior works aim to learn various emotions at once, leading to inadequate modeling. 
This paper explores the disentanglement of emotions in piano performance generation through a two-stage framework. 
The first stage focuses on valence modeling of lead sheet, and the second stage addresses arousal modeling by introducing performance-level attributes. 
To further capture features that shape  valence, an aspect less explored by previous approaches, we introduce a novel functional representation of symbolic music.
This representation aims to capture the emotional impact of major-minor tonality, as well as the interactions among notes, chords, and key signatures.
Objective and subjective experiments validate the effectiveness of our framework in both emotional valence and arousal modeling. 
We further leverage our framework in a novel application of emotional controls, showing a broad potential in emotion-driven music generation.
\end{abstract}
\section{Introduction}\label{sec:introduction}
With the recent advancements in symbolic music generation~\cite{musictransformer, remi, figaro, SketchNet,wu2021musemorphose,musecoco}, there has been a growing interest in controlling high-level musical features throughout the generation process.
Among these features, \emph{emotion-driven music generation}~\cite{emopia, transformer-GAN, muser, emogen, learningto, YM2413-MDB, MoodLoopGP} aims to generate music that conveys specific emotions, representing a crucial aspect for music appreciation and analysis. 
The downstream applications of such models have also been explored, such as music therapy for healthcare and educational purposes~\cite{fpsyg22} and soundtrack generation for videos and movies~\cite{10447950}.

Emotion could be represented in two dimensions from the literature~\cite{russell}: \emph{valence} and \emph{arousal}.
Valence refers to the positiveness of an emotion and arousal refers to energy or activation~\cite{regression-based,emotion-rev,mer}.
These two dimensions can be further divided into four quadrants (4Q), namely high valence high arousal (Q1),  low valence high arousal (Q2), low valence low arousal (Q3), and high valence low arousal (Q4). 
In this paper, we focus on the emotion-driven \emph{piano performance generation} of these four quadrants. 
Throughout prior works, we observe crucial challenges from the perspectives of both model design and musical inductive bias.

First, previous emotion-driven piano performance generation models~\cite{emopia, transformer-GAN} attempt to learn emotion quadrants and expressions in an \emph{end-to-end} paradigm. 
In terms of model design, this approach poses training difficulty on the generation model, leading to the instability in achieving results of desired emotions. 
For example, many existing works~\cite{emopia, muser, learningto} could effectively control the arousal levels of music, while their performance of \emph{valence modeling}, especially in generating low valence (i.e., negative) music, is still poor. 
In terms of music, the creation process of music typically involves multiple stages, such as the \emph{lead sheet composition} for melodies and chord progressions, and \emph{performance generation} for textures and expressiveness. 
Consistently, emotion can be evoked through a combination of musical elements (e.g., melody, chord, texture). 
For example, major/minor chords have been found to seize different valence trends in psychological studies~\cite{bakker15} and performance-level attributes like articulation, tempo, and velocity are more related to arousal~\cite{generationof, onperceived}. 
It is worth to explore the potential relation between the \emph{disentanglement} of the generation process and the emotion expression. 
%
%
%
%
%

\begin{figure}[t]
 \centerline{
 \includegraphics[width=\columnwidth]{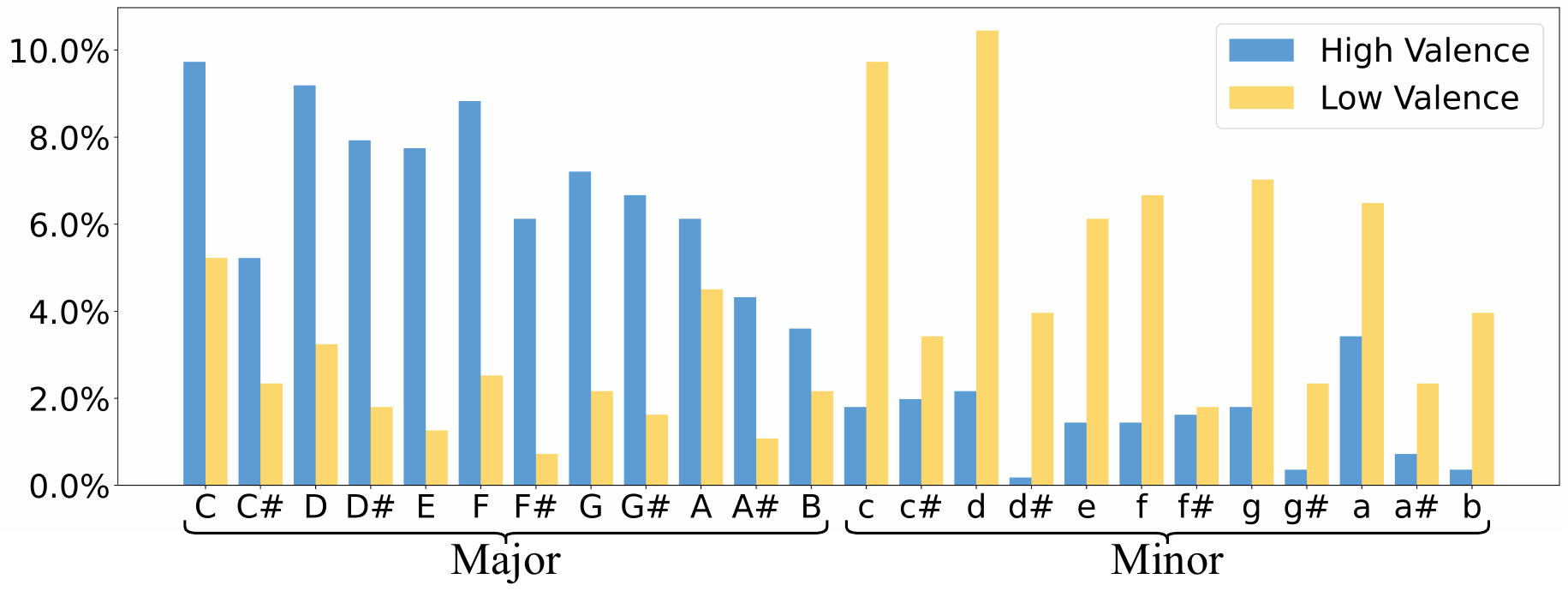}}
 \vspace{-0.2cm}
 \caption{Key histogram of high/low valence clips from the emotion-labeled piano music dataset EMOPIA \cite{emopia}.}
 \label{fig:key}
\end{figure}

Second, previous emotion-driven generation models have received limited attention regarding the influence of \emph{tonality} on emotion modeling.
It has been widely shown that major-minor tonality in composition is highly related to valence perception~\cite{audiofeatures, onperceived, adata-driven, TheEffects}. 
For example, as depicted in Figure~\ref{fig:key}, the histogram of musical keys derived from the emotion-labeled music dataset EMOPIA~\cite{emopia} supports the distribution skews to major keys for high valence clips and opposite trend for low valence ones. 
Furthermore, different tonalities may reveal similar patterns in the relative relationships between melodies and chords, while the distribution of melodies, chords, and tonalities can exhibit distinct shapes across different emotions.
Current representations of symbolic music, such as REMI~\cite{remi} and CP-Word~\cite{cp}, do not explicitly incorporate such interactions nor address its connection to emotion adequately.
Therefore, it is necessary to consider a functional format of symbolic music representation considering the relationships between notes, chords and key signatures to better model the tonality in the emotion-driven music generation process.

In this paper, we contribute to combat above challenges:
\begin{itemize}[leftmargin=*,itemsep=0pt,topsep=2pt]
    \item We employ a two-stage Transformer-based model on emotion-driven piano performance generation. The first stage focuses on valence modeling via lead sheet composition, while the second stage addresses arousal modeling by introducing performance-level attributes.
    
    \item We propose a novel functional representation for symbolic music, encoding both melody and chords with Roman numerals relative to musical keys, to consider the interactions among notes, chords and tonalities~\cite{functionalharmony}.
    
    \item Experiments demonstrate the effectiveness of our framework and representation on emotion modeling. Additionally, our method enables new capabilities to control the arousal levels of generation under the same lead sheet, leading to more flexible emotion controls.
\end{itemize}
As a minor contribution, we also refine key signature labels and extract lead sheet annotations for the EMOPIA dataset~\cite{emopia} to ensure the correct training of the two-stage framework. 
We share the data, open source our code\footnote{\url{https://github.com/Yuer867/EMO-Disentanger}} and present generation samples in the demo page.\footnote{\url{https://emo-disentanger.github.io/}}

\section{Related Work}\label{sec:related_work}
\subsection{Emotion-driven Piano Performance Generation} 
Prior works apply emotion conditions on deep-learning models to guide the generation of piano performance~\cite{learningto, emopia, transformer-GAN}, or develop searching methods to generate music of desired emotions~\cite{Controlling, Computer-Generated}.
Musical elements via feature disentanglement~\cite{muser} or supervised clustering~\cite{emogen} can further be regarded as a bridge between emotion labels and performances for generation.
In contrast, our framework employs a two-stage generation approach to reduce the complexities of one-stage generation, fostering a more nature process of music creation as well as a better incorporation between emotion labels and generation results.

\subsection{Tonality, Functional Harmony, and Emotion}

Musical keys and functional harmony have been explored in the field of roman numeral analysis~\cite{AugmentedNet, notallroads, ChordGNN}. 
The analysis of how modes and tonalities relate to mid-level perceptual features (e.g., dissonance, tonal stability, minorness) and affect the emotional perception of music pieces has also been discovered~\cite{onperceived, adata-driven}.

While some music generation works attempted to combine key information into data representation~\cite{Musictranscription}, loss function~\cite{mtharmonizer} and text conditions~\cite{musecoco}, none of them explore the relation between musical keys and emotional perception. 
In this paper, we leverage both functional harmony knowledge and class-octave based pitch representation~\cite{class-octave} to design a new data representation, incorporating the relationships between notes, chords and keys for emotion-driven music generation.

\section{Method}\label{sec:method}
In this section, we will first introduce the functional representation of symbolic music as the main generation unit.
Then we introduce the two-stage model as the main component of the emotion disentanglement and generation. 
\subsection{Functional Representation}\label{subsec:functional}

\begin{figure}[t]
    \centering
    \includegraphics[width=\columnwidth]{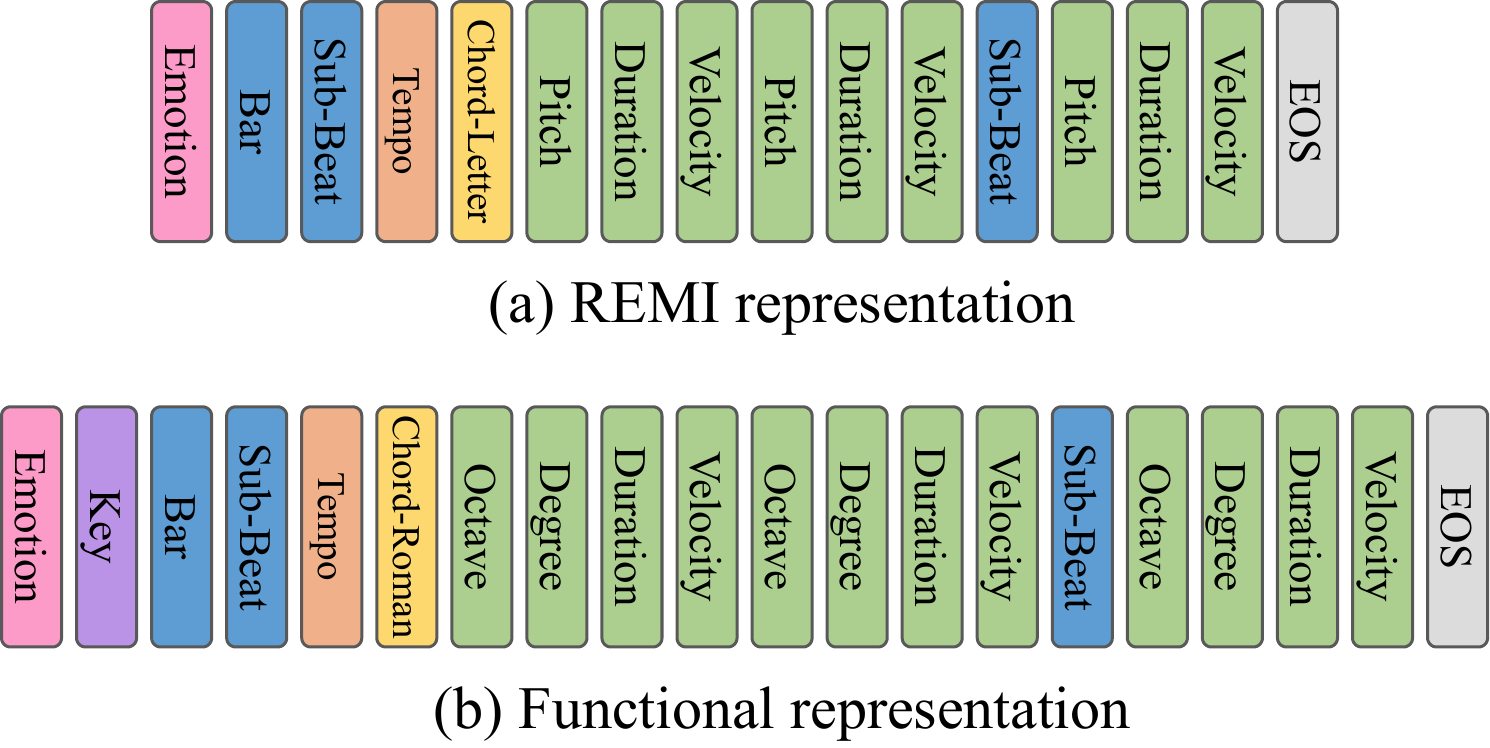}
    \caption{Illustration of (a) REMI \cite{remi}, (b) the proposed functional representation, and their differences.}
    \label{fig:representation}
\end{figure}

Figure~\ref{fig:representation} illustrates our proposed functional representation.
Its design is initially based on REMI~\cite{remi}, a widely used event-based representation for symbolic music. 
We incorporate different note and chord events assisting to better learn the joint information of emotion and key signature.

\subsubsection{Emotion and Key Events}
We follow CTRL~\cite{CTRL} to set up the condition within the autoregressive generation process in Transformer architecture.
To denote distinct emotions and affect overall properties, we begin the event sequence with \texttt{<Emotion\_*>} event to indicate the emotion label of music clips. 
The \texttt{<Key\_*>} event is appended after \texttt{<Emotion\_*>} to provide the musical key property, with the total of 24 keys (12 tonic notes with two modes in EMOPIA~\cite{emopia}). 

\subsubsection{Bar, Sub-Beat, Tempo and EOS Events}
Similar to REMI, 
a \texttt{<Bar>} event denotes the new start of a bar; 
a \texttt{<Sub-Beat\_*>} event denotes one of 16 possible discrete beat locations within a bar;
a \texttt{<Tempo\_*>} event denotes local tempo changes every four beats;
and an \texttt{<EOS>} event denotes the end of sequence. 

\subsubsection{Chord Events}
A musical chord name typically consists of root note and chord quality. 
For example, \texttt{Fmaj} represents the chord \texttt{F-A-C} with root \texttt{F} and major quality.
%
Such symbols describe correct note information in chord within the tonality, but they overlook the variations in \emph{chord functions} of the same chord across different tonalities.
For example, while \texttt{Fmaj} serves the tonic function in F major scale, it serves the subdominant function in C major scale. 
Moreover, the chord progression follows these functional harmony rules to establish tonality and convey musical emotion~\cite{functionalharmony}.

\begin{figure}
    \centering
    \includegraphics[width=\columnwidth]{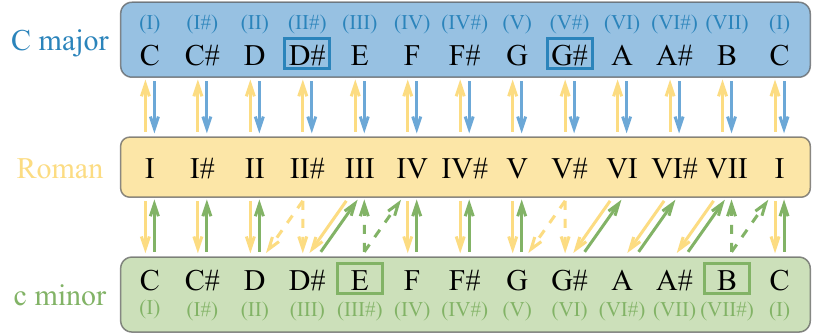}
    \caption{The conversion between letters and Roman numerals in the cases of C major and c minor scales. Solid arrows denote strict one-to-one conversions, and dotted arrows denote optional one-to-either conversions.}
    \label{fig:switch}
\end{figure}

To introduce chord functions in the emotion modeling, we adopt Roman numerals from \textit{Roman Numeral Analysis}~\cite{notallroads} to notate chord roots in Figure~\ref{fig:switch}.
Given the \texttt{<Key\_*>} event, root notes in the absolute pitch are directly converted into Roman numerals based on their scale degrees relative to the key (i.e., relative pitch). 
For roots outside the scale, we employ a direct conversion for \texttt{I\#}, \texttt{II\#}, \texttt{IV\#}, \texttt{V\#} and \texttt{VI\#} appearing in major keys, but randomly assign \texttt{III\#} and \texttt{VII\#}, which only appear in minor keys, as one of their neighboring degrees during the encoding and decoding process. 
This design ensures the notation to be key-independent and make every conversion of notes reasonable to the music theory. 
The notations of chord qualities remain unchanged, and the chord event \texttt{<Chord\_*>} appears every four beats.

\subsubsection{Note-related Events} 
\begin{figure}[t]
    \centering
    \includegraphics[width=\columnwidth]{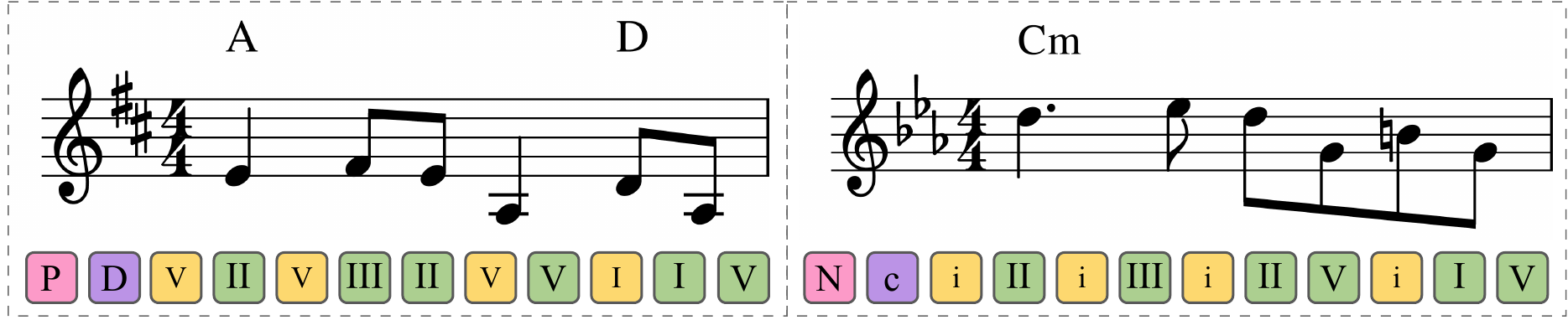}
    \caption{Two lead sheet examples from different songs in EMOPIA. In our functional representation, they have the same melody events (green), but different chord events (yellow) by different emotions (\underline{P}ositive and \underline{N}egative by pink) and keys (\underline{D} major or \underline{c} minor by purple).}
    \label{fig:samples}
    \vspace{-0.3cm}
\end{figure}

A note is denoted by \texttt{<Pitch\_*>}, \texttt{<Duration\_*>} and \texttt{<Velocity\_*>} events, where \texttt{<Pitch\_*>} event indicates the onset of pitches from A0 to C8. 
Inspired by~\cite{class-octave, musicsimilarity}, we decompose \texttt{<Pitch\_*>} into \texttt{<Octave\_*>} and \texttt{<Degree\_*>} events according to the note octave and degree in the certain key scale. 
The conversion rule from \texttt{<Pitch\_*>} to  \texttt{<Degree\_*>} is the same as that of chord roots in Figure \ref{fig:switch}.
For example, pitch D\#4 is decomposed into \texttt{<Octave\_4>} and \texttt{<Degree\_III>} in c minor scale, but \texttt{<Degree\_I>} in D\# major scale.
Such degree-octave pitch representation narrows the difference between melodies, thus improves the learning of connections between emotions, chords, and melodies, as demonstrated in Figure~\ref{fig:samples}.

\subsection{Two-stage Emotion Disentanglement}\label{subsec:two_stage}

We use the idea of Compose \& Embellish~\cite{c&e} to generate music in two stages: lead sheet first, and then piano performance. While Compose \& Embellish is emotion-agnostic, we extend it so that the lead sheet model involves \emph{valence modeling} and the performance model  \emph{arousal modeling}. 

\begin{figure*}
    \centering
    \includegraphics[width=0.9\textwidth]{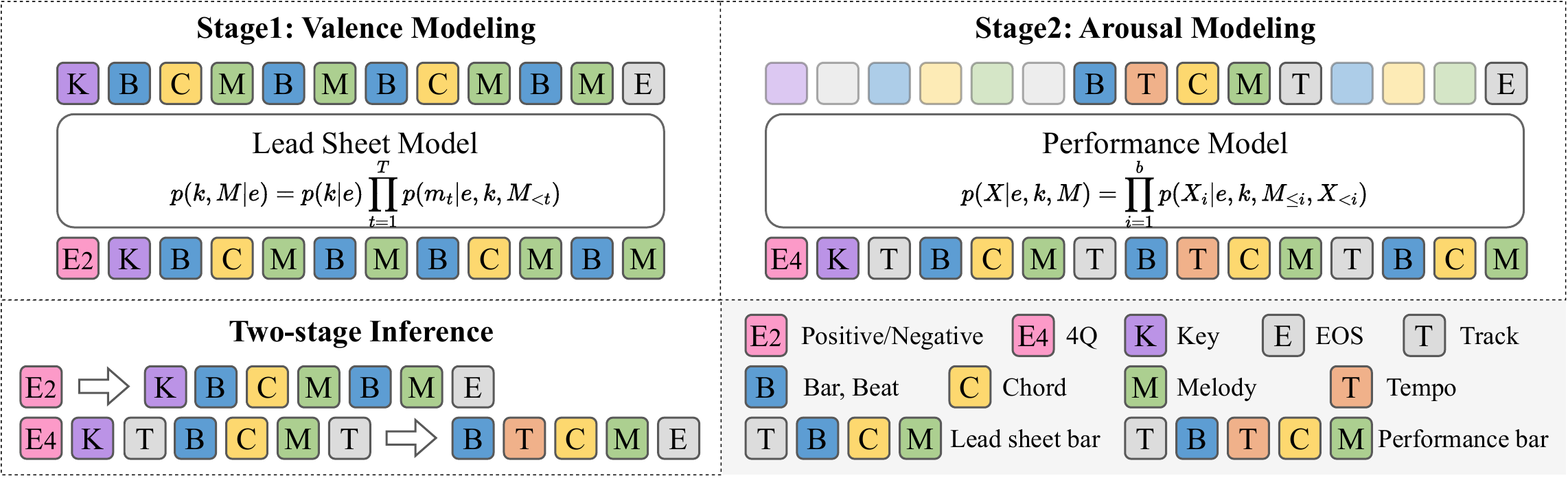}
    \caption{The two-stage framework of emotion-driven piano performance generation. Squares with transparent background denote the tokens that are not included in the loss computation during the training phase.}
    \label{fig:model}
    \vspace{-0.1cm}
\end{figure*}

\subsubsection{Valence Modeling}
The top left section of Figure~\ref{fig:model} denotes the first stage, where only emotion events \texttt{<Emotion\_Positive>} and \texttt{<Emotion\_Negative>} are considered as conditions.
The former includes music pieces of Q1 and Q4 (high valence) and the latter includes those of Q2 and Q3 (low valence).
The lead sheet model first predicts a key event $k$ conditioned on the given emotion event $e$, and then generates the lead sheet sequence $M=\{m_1, \cdots m_T\}$ of length $T$, as melody and chord progression, conditioned on previous tokens step-by-step:
\begin{equation}
    p(k, M|e) 
    = p(k|e)\prod_{t=1}^T p(m_t | e, k, M_{<t}) \,,
\end{equation}
where $p(k|e)$ and $p(m_t | e, k, M_{<t})$ are jointly learned through the Transformer-based generation model~\cite{cp, c&e}. 
Performance-related events \texttt{<Velocity\_*>} and \texttt{<Tempo\_*>} are removed in the first stage (i.e., lead sheet generation), as we mainly focus on the contributions of key, pitch and chord for valence perception.

\subsubsection{Arousal Modeling}
The top right section of Figure~\ref{fig:model} denotes the second stage.
Given the lead sheet $M$, the performance model generates performance $X$ conditioned on the true emotion label (Q1 to Q4). 
As the valence aspect has already been modeling in the first stage, this stage focuses on the generation of musical textures for the lead sheet, and more importantly, on how to perform it through variations of tempo, velocity, articulation, and other performance-level attributes that largely influence perceived arousal~\cite{onperceived, generationof}.
During the training and inference phases, with the positions of \texttt{<Bar>} events, $M$ and $X$ are further segmented into $\{M_1, \cdots, M_b\}$ and $\{X_1, \cdots, X_b\}$, where $b$ is the number of bars. 
The segmented sequences are ``interleaved'' in the form of $\{ \cdots $\texttt{<Track\_M>}, $M_i$, \texttt{<Track\_X>}, $X_i \cdots \}$ with additional \texttt{<Track\_*>} events to distinguish $M$ and $X$ tracks.
In that, the target performance bar $X_i$ is appended to its corresponding conditions $M_i$, as mapping each lead sheet segment to its corresponding performance segment~\cite{cp}.
With the emotion condition and key event from lead sheet as prefix tokens, the performance model is summarized as
\begin{equation}
    p(X|e, k, M) = \prod_{i=1}^b p(X_i|e, k, M_{\leq i}, X_{<i}) \,.
\end{equation}

\subsubsection{Training Objectives}\label{sec:training_obj}
Lead sheet and performance models are trained separately by both optimizing the negative log-likelihood loss of the sequence.
Since existing emotion-labeled music datasets are not large, we leverage large-scale music datasets without emotion annotations to \emph{pretrain} both models for better music understanding.
During pretraining, the emotion event is marked as \texttt{<Emotion\_None>}.
We then finetune two models on the emotion-labeled dataset (detail in Section~\ref{sec:experiments})
to learn composition and performance styles specific to different emotion contexts.

\subsubsection{Two-stage Inference} 
The left bottom section of Figure~\ref{fig:model} denotes the inference process of both models.
In the first stage, the lead sheet model predicts the key event and generates the lead sheet sequence step-by-step given \texttt{<Emotion\_Positive>} or \texttt{<Emotion\_Negative>} event, creating a musical motif for the specific valence preference.
Even though our framework has the capability to generate any-key music of specific emotions, we observe that some generation results, such as a high-valence and high-arousal song with a minor key scale, may go beyond the current definition of emotion in~\cite{russell}, where the valence naturally has a strong correlation to the major-minor tonality (Figure~\ref{fig:key}).
Therefore, we limit major keys to \texttt{<Emotion\_Positive>} and minor keys to \texttt{<Emotion\_Negative>} \emph{during the inference stage}. 
We acknowledge that this can be overly simplifying.
Since this paper focuses mainly on the valence-arousal disentanglement during the generation process, we leave this exploration of generating any emotion within any key as an advanced topic for future research. 

In the second stage, the performance model generates piano performance with desired valence and arousal combination given the lead sheet from the first stage. 
For example, to generate a music piece of Q3, a ``Negative'' lead sheet and a ``Q3'' emotion event are selected as conditions.
Additionally, this two-stage framework enables the flexibility to generate different arousal levels of piano performance under the same lead sheet, delivering some scenarios when the music need to shift quickly to complement the scenes in movies or daily videos (detail in Section~\ref{sec:experiments} and the demo page). 

\section{Experiments}\label{sec:experiments}

\subsection{Datasets and Preprocessing}\label{subsec:dataset}
As presented in Table~\ref{table:dataset}, we collect different datasets for pretraining and finetuning phases as mentioned in Section ~\ref{sec:training_obj}. 
For pretraining the lead sheet model, we use the HookTheory dataset~\cite{hooktheory, sheetsage}, where we choose 18,206 lead sheets with high-quality and human-transcribed melody, chord and key annotations in 4/4 time signature. 
%
We simplify 249 chord quality classes into 11 types\footnote{Major, minor, augment, diminish, suspend2, suspend4, major7, minor7, dominant7, diminish7, half-diminish7} as the same set in the other datasets below.
For pretraining the performance model, we use the Pop1k7 dataset~\cite{cp}, consisting of 1747 transcribed pop piano performances.
Since Pop1k7 does not contain lead sheet annotations, we refer \cite{c&e} to extract melodies using the skyline algorithm~\cite{skyline}, recognize chords using the chorder library~\cite{chorder}, and detect key signatures using \cite{KS-key} in MIDI Toolbox\cite{miditoolbox2016}.

For finetuning the models with emotion conditions, we use the EMOPIA dataset~\cite{emopia}, consisting of 1,071 music clips with human-annotated emotion labels. 
Similar to Pop1k7, we obtain the lead sheets of EMOPIA by extracting melodies using the algorithm in \cite{miditoolkit} and recognizing chords using the algorithm in\cite{pop909}.
%
Empirically, we obverse that specifically in the EMOPIA dataset, melodies and chords extracted by these alternative algorithms are more correct compared to the skyline algorithm and the chorder library. 
%
Additionally, we found the key signature labels in EMOPIA are not fully correct since they are also obtained by the detection algorithm with error rates.
%
Since the valence modeling is strongly related to the musical keys and modes, we manually correct the key annotations of 367 clips in EMOPIA to ensure a high quality of lead sheets.

All datasets are randomly divided into respective training and validation sets at the ratio of 9:1. 
As a result in our functional representation, the vocabulary size of events is 215 for lead sheet and 324 for piano performance.


\begin{table}[t]
\centering
\resizebox{\columnwidth}{!}{
\begin{tabular}{lrrr}
\toprule
\textbf{Dataset} & \textbf{\# clips ~(major)} & \textbf{\# bars} & \textbf{\# events}  \\
\midrule
HookTheory\,\cite{sheetsage}        & 18,206 ~(9,737)            & 10.84           & 282.81          \\
Pop1k7\,\cite{cp}                   & 1,747 ~(1,264)              & 104.82          & 6794.86         \\
EMOPIA(L)\,\cite{emopia}            & 1,071 ~~~~(618)               & 16.94           & 435.22          \\
EMOPIA\,\cite{emopia}               & 1,071 ~~~~(618)               & 17.09           & 1311.47         \\
\bottomrule
\end{tabular}
}
\caption{The datasets. (major) denotes the number of clips in major key (and the left is in minor key). The \#bars and \#events are average numbers across a dataset. EMOPIA(L) refers to EMOPIA lead sheets. }
\label{table:dataset}
\vspace{-0.1cm}
\end{table}

\subsection{Model Settings}
The lead sheet model is a 12-layer Transformer Decoder~\cite{transformer} with 8 heads, 512 hidden dimensions and relative positional encoding~\cite{Transformer-XL}. 
The performance model is similar to the lead sheet model except with Performer attention~\cite{performer}.
The total parameter sizes are 41 million and 38 million respectively.

Both models are trained with the batch size of 4, the maximum sequence length of 512 (lead sheet model) or 3072 (performance model), and the Adam optimizer with $\beta=(0.99,0.9)$.
We adopt a 200-step warm-up to achieve the maximum learning rate of 1e-4 for pretraining and 1e-5 for finetune.
All models are implemented by PyTorch and trained on one NVIDIA Tesla V100 GPU. 
The lead sheet model took around 180,000 steps to converge and the performance model took around 200,000 steps.
Nucleus sampling~\cite{sampling} is employed in the inference phase.
We referred \cite{ATheoretical, c&e} to choose the sampling hyperparameters $\tau=1.2$, $p=0.97$ for the lead sheet model and $\tau=1.1$, $p=0.99$ for the performance model.

\subsection{Baseline and Ablations}
We consider the emotion-driven piano performance generation model in EMOPIA~\cite{emopia} as our baseline, which generates music in an end-to-end paradigm instead of two stages. 
To ensure the fair comparison of generation performance, we trained the baseline model under ths same datasets in both pretraining and finetune phases, and replaced the original CP-Word representation with REMI as the former usually yields better generation performance and more comparable to our proposed functional representation.  
%
Two other related works~\cite{muser, transformer-GAN} are not included in comparison due to the main reason that we focus more on the evaluation of the two-stage framework in valence and arousal modeling; and the partial reason that they are not open-source or releasing the reproducible model weights.

We conduct a comprehensive ablation study to evaluate if each proposed design benefits the emotion modeling of music generation. 
Specially, these designs include: 1) the two-stage generation, 2) the functional representation, and 3) the dataset pretraining.
%
In the following sections, models are denoted as \textbf{<representation(stage)>}. For example, REMI(one) denotes the one-stage generation model with REMI representation as the baseline, and REMI(two) denotes the two-stage generation as one variant.

\begin{table}[t]
\centering
\resizebox{0.85\columnwidth}{!}{
\begin{tabular}{lcccc}
\toprule
    & \textbf{M} & \textbf{C}  & \textbf{M+C} & \textbf{P}  \\
\midrule
REMI+key\,(two)       &  0.465        & 0.065          & 0.075        &  0.418              \\
\quad --w/o. pretrain  &  0.350        & 0.105          & 0.130        &  0.343              \\
functional\,(two)     &  \textbf{0.505}   & \textbf{0.700}   & \textbf{0.735} &  \textbf{0.548}              \\
\quad --w/o. pretrain  &  0.400        & 0.570          & 0.625        &  0.430              \\
\midrule
Real data             &  0.578        & 0.695          & 0.746        &  0.812              \\
\bottomrule
\end{tabular}
}
\caption{Key consistency calculated across all components, including melody (\textbf{M}), chord (\textbf{C}), lead sheet (\textbf{M+C}) and performance (\textbf{P}).}
\label{table:key}
\vspace{-0.1cm}
\end{table}

\subsection{Objective Evaluation and Results}\label{subsec:objective}
Even though previous studies~\cite{emopia, transformer-GAN, muser} employ metrics, such as Pitch Range (PR) and Number of Pitch Classes (NPC), to evaluate the generation performance, they do not provide any evidences on the superiority of  melody development, chord progression, and texture arrangement of music. 
Therefore, a model with more similar PR and NPC values to those of the target dataset does not necessarily promise a better generation quality than others.

Instead of using such metrics, we wish to evaluate the consistency between the input musical conditions and the generation results. 
We introduce \textbf{key consistency} to assess if a model can generate music pieces that adhere to the desired input key signatures, which is highly correlated to the lead sheet development and valence modeling. 
Specifically, key consistency measures the match between the key condition \texttt{<Key\_*>} and the actual key detected in the generation using the algorithm~\cite{miditoolbox2016} with an 81\% accuracy rate. 
We compare REMI and our functional representation to determine if the functional representation can improve the key consistency via more close and interactive designs on key, melody, and chord. 
Since this metric requires the key as conditions, we add the \texttt{<Key\_*>} event in REMI after \texttt{<Emotion\_*>} (as REMI+key in Table~\ref{table:key}) when training the model. 
The non-pretrained versions are also included for comparison.
Each model generates 200 lead sheets (100 high and 100 low valence) and 400 performance samples (100 per emotion quadrant) for evaluation.

From Table~\ref{table:key}, the functional representation outperforms REMI (i.e., REMI+key) across all components and even achieves compatible accuracy to real data over the lead sheet component.
This demonstrates the effectiveness of the functional representation, by representing notes and chord roots relative to key events for key modeling.
In contrast, REMI struggles with associating chord events with keys due to the ignorance of chord labels serving different functions in different key scales. 
Moreover, pretraining process introduces musical priors to enhance the learning of key relationships with other musical elements, improving key consistency for both representations.


\subsection{Subjective Evaluation and Results}\label{subsec:subjective}
We leverage an online listening test to assess the emotion modeling ability of models. 
The test was conducted to collect user responses on three parts: 1) valence modeling, 2) arousal modeling, and 3) 4Q emotion modeling. 
During this test, the quality of the generated music has also been assessed implicitly as it is a prerequisite to the emotion expression in the music.
22 participants were engaging in this test, 5 with less than 2 years of musical training, 8 with 2-5 years, 3 with 5-10 years, and 6 with more than 10 years.

\subsubsection{Valence Modeling}
In this part, each participant listened to 16 generated tracks of piano performance from four models [four tracks (two high valence and two low valence) per model]: 1) Real data; 2) REMI (one); 3) REMI (two); 4) Functional (two).
%
For each track, participants rated its positiveness from --2 (low valence) to 2 (high valence) with the step size 1.

\begin{figure}[t] 
	\centering	
    \includegraphics[width=\columnwidth]{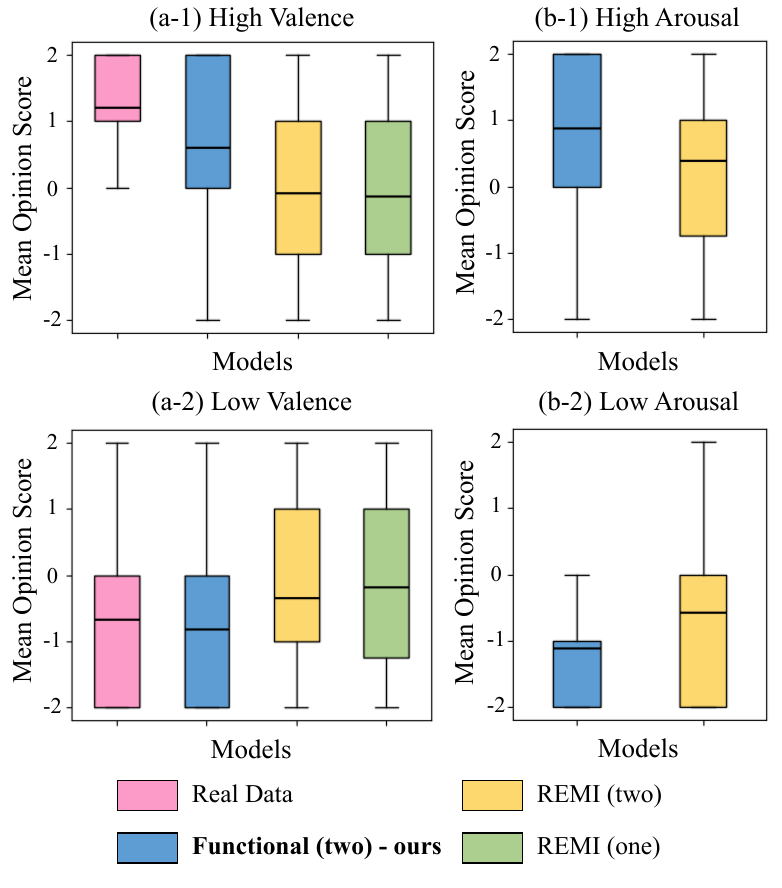}
    \vspace{-0.8cm}
	\caption{The mean opinion score performance on the valence-oriented and arousal-oriented listening tests. For (a-1) and (b-1), the higher score the better performance; for (a-2) and (b-2), the lower score the better performance.}
    \label{fig:2Q}
    \vspace{-0.1cm}
\end{figure}

The left of Figure~\ref{fig:2Q} (`(a)') presents the mean opinion scores for the valence-oriented test, where the Functional (two) model significantly outperforms both REMI (two) and REMI (one) models.
The REMI (two) model shows a slight improvement over REMI (one) due to its two-stage design.
Our proposed Functional (two) model even marginally exceeds real data in low valence scores, which could be due to the potential subjective biases in the negative emotion as discussed in~\cite{emogen}. 
And our model achieves both great performance in high valence and low valence results, demonstrating a good balance in valence modeling.

\subsubsection{Arousal Modeling}\label{subsec:arousal}
In the second part, the functional (two) and REMI (two) models are chosen to compare their arousal modeling performance.
Specifically, we wish to explore whether they can generate piano performance with either high or low arousal under the same lead sheet based on the given conditions (Q1 and Q4 for positive lead sheets, Q2 and Q3 for negative ones). 
Two pairs of generated tracks are randomly drawn for each model and each valence level, where every pair includes two tracks of different arousal conditions.
For each track, participants rated its arousal level from --2 (low arousal) to 2 (high arousal) with the step size 1. 

The right of Figure~\ref{fig:2Q}  (`(b)') presents the results. 
The Functional (two) model surpasses REMI (two) by an average of 0.5 point, highlighting its superior ability to differentiate between the musical features of the two arousal levels through performance. 
Additionally, it is rare for Functional (two) to be incorrectly identified as high arousal under low arousal conditions (Figure \ref{fig:2Q} (b-2)).

\subsubsection{4Q Emotion Judgement}
In the last part, participates needed to choose the best option from four options (4Q) for each track, with 8 tracks in total for the two models the last section (4 tracks per model and 1 track per emotion).

\begin{figure}[t] 
	\centering	
    \includegraphics[width=\columnwidth]{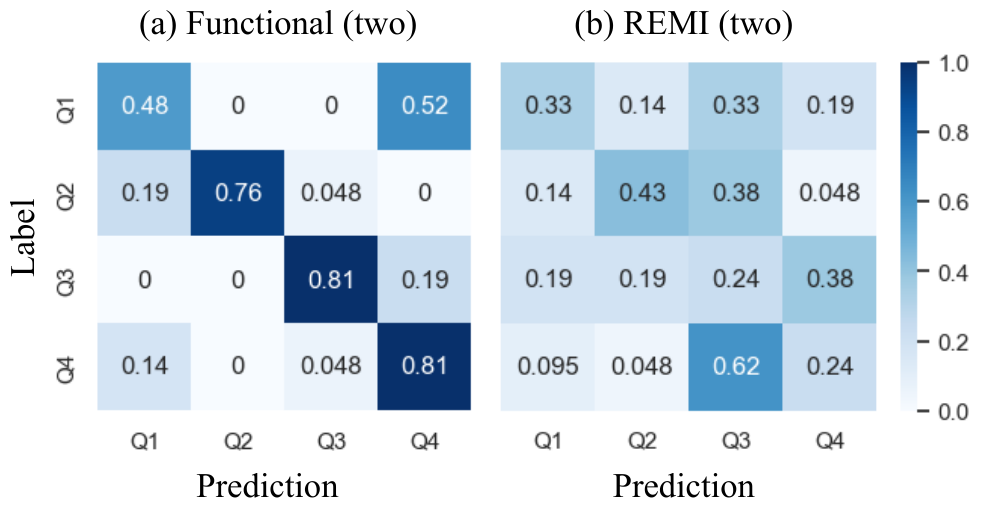}
	\caption{The confusion matrices on the 4Q listening tests.}
    \label{fig:4Q}
    \vspace{-0.3cm}
\end{figure}

Figure~\ref{fig:4Q} presents the confusion matrices of two models.
The Functional (two) model achieves the higher overall accuracy than that of REMI (two) (71.5\% vs. 31.0\%).
When examining each emotion category, Functional(two) demonstrates superior performance in Q3 and Q4 than Q1 and Q2. 
Furthermore, music pieces generated from it with high valence conditions are misidentified almost based on their arousal levels; for instance, pieces intended for Q1 are almost mistaken for Q4 and vice versa. 
In contrast, for REMI(two), the misclassifications are across all categories, demonstrating its limitations in modeling the four emotion classes although through two-stage generation.

All above evaluations support that the combination of two-stage framework and functional representation is effective in controlling the emotion of the music it generates to a certain extent.

\section{Conclusion and Future Work}
In this paper, we first explore emotion disentanglement through a two-stage Transformer-based framework for emotion-driven piano performance generation. Then we propose a novel functional representation for symbolic music to capture the interactions among musical keys, modes, chords, and melodies in relation to the emotion contexts. 
An objective metric is designed to qualify the key modeling of the proposed method, and subjective evaluations further confirm its ability to convey desired emotional perception. 
In the future, we wish to focus on enhancing the flexibility of emotional music generation across all musical keys and investigating new applications fostered by our framework, such as the controls of valence and arousal attributes under similar music motifs.

\section{Acknowledgment}

The work is supported by a grant from the National Science and Technology Council of Taiwan (NSTC 112-2222-E-002-005-MY2). This work also benefits from the review comments of an unpublished work \cite{Emo-Harmonizer}, which is an early version of the present study. While the concept of functional representation has been mentioned in that previous work \cite{Emo-Harmonizer}, its focus was on emotion-driven melody harmonization and the effectiveness of the functional representation in that context was unclear, as the melody was pre-given and fixed. In the current work, melody and chords are generated together, allowing more flexibility for emotion control with the proposed functional representation. The source code of that previous work can be found at the GitHub repo: \url{https://github.com/Yuer867/EMO_Harmonizer}.

\bibliography{refs}

\begin{thebibliography}{10}
\providecommand{\url}[1]{#1}
\csname url@samestyle\endcsname
\providecommand{\newblock}{\relax}
\providecommand{\bibinfo}[2]{#2}
\providecommand{\BIBentrySTDinterwordspacing}{\spaceskip=0pt\relax}
\providecommand{\BIBentryALTinterwordstretchfactor}{4}
\providecommand{\BIBentryALTinterwordspacing}{\spaceskip=\fontdimen2\font plus
\BIBentryALTinterwordstretchfactor\fontdimen3\font minus \fontdimen4\font\relax}
\providecommand{\BIBforeignlanguage}[2]{{%
\expandafter\ifx\csname l@#1\endcsname\relax
\typeout{** WARNING: IEEEtran.bst: No hyphenation pattern has been}%
\typeout{** loaded for the language `#1'. Using the pattern for}%
\typeout{** the default language instead.}%
\else
\language=\csname l@#1\endcsname
\fi
#2}}
\providecommand{\BIBdecl}{\relax}
\BIBdecl

\bibitem{musictransformer}
C.-Z.~A. Huang, A.~Vaswani, J.~Uszkoreit, I.~Simon, C.~Hawthorne, N.~Shazeer, A.~M. Dai, M.~D. Hoffman, M.~Dinculescu, and D.~Eck, ``{Music Transformer}: Generating music with long-term structure,'' in \emph{Proc. ICLR}, 2019.

\bibitem{remi}
Y.-S. Huang and Y.-H. Yang, ``{Pop Music Transformer}: Beat-based modeling and generation of expressive pop piano compositions,'' in \emph{Proc. ACM Multimed.}, 2020.

\bibitem{figaro}
D.~von Rütte, L.~Biggio, Y.~Kilcher, and T.~Hofmann, ``{FIGARO}: Generating symbolic music with fine-grained artistic control,'' in \emph{Proc. ICLR}, 2023.

\bibitem{SketchNet}
K.~Chen, C.~Wang, T.~Berg{-}Kirkpatrick, and S.~Dubnov, ``Music sketchnet: Controllable music generation via factorized representations of pitch and rhythm,'' in \emph{Proc. ISMIR}, 2020.

\bibitem{wu2021musemorphose}
S.-L. Wu and Y.-H. Yang, ``{MuseMorphose}: Full-song and fine-grained piano music style transfer with one {Transformer VAE},'' \emph{IEEE Trans. Audio, Speech, Lang. Process.}, vol.~31, pp. 1953--1967, 2023.

\bibitem{musecoco}
P.~Lu, X.~Xu, C.~Kang, B.~Yu, C.~Xing, X.~Tan, and J.~Bian, ``{MuseCoco}: Generating symbolic music from text,'' \emph{CoRR}, vol. abs/2306.00110, 2023.

\bibitem{emopia}
H.~Hung, J.~Ching, S.~Doh, N.~Kim, J.~Nam, and Y.-H. Yang, ``{EMOPIA:} {A} multi-modal pop piano dataset for emotion recognition and emotion-based music generation,'' in \emph{Proc. ISMIR}, 2021.

\bibitem{transformer-GAN}
P.~L.~T. Neves, J.~Fornari, and J.~B. Florindo, ``Generating music with sentiment using {T}ransformer-{GAN}s,'' in \emph{Proc. ISMIR}, 2022.

\bibitem{muser}
S.~Ji and X.~Yang, ``Mus{ER}: Musical element-based regularization for generating symbolic music with emotion,'' in \emph{Proc. AAAI}, 2024.

\bibitem{emogen}
C.~Kang, P.~Lu, B.~Yu, X.~Tan, W.~Ye, S.~Zhang, and J.~Bian, ``{EmoGen}: Eliminating subjective bias in emotional music generation,'' \emph{CoRR}, vol. abs/2307.01229, 2023.

\bibitem{learningto}
L.~Ferreira and J.~Whitehead, ``Learning to generate music with sentiment,'' in \emph{Proc. ISMIR}, 2019.

\bibitem{YM2413-MDB}
E.~Choi, Y.~Chung, S.~Lee, J.~Jeon, T.~Kwon, and J.~Nam, ``{YM2413-MDB}: A multi-instrumental {FM} video game music dataset with emotion annotations,'' in \emph{Proc. ISMIR}, 2022.

\bibitem{MoodLoopGP}
W.~Cui, P.~Sarmento, and M.~Barthet, ``Mood{L}oop{GP}: Generating emotion-conditioned loop tablature music with multi-granular features,'' in \emph{Proc. EvoMUSART}, 2024.

\bibitem{fpsyg22}
K.~Zheng, R.~Meng, C.~Zheng, X.~Li, J.~Sang, J.~Cai, J.~Wang, and X.~Wang, ``{EmotionBox}: A music-element-driven emotional music generation system based on music psychology,'' \emph{Frontiers in Psychology}, vol.~13, 2022.

\bibitem{10447950}
M.~T. Haseeb, A.~Hammoudeh, and G.~Xia, ``{GPT-4} driven cinematic music generation through text processing,'' in \emph{Proc. ICASSP}, 2024.

\bibitem{russell}
J.~A. Russell, ``A circumplex model of affect,'' \emph{Journal of Personality and Social Psychology}, 1980.

\bibitem{regression-based}
Y.-H. Yang, Y.-C. Lin, Y.-F. Su, and H.~H. Chen, ``A regression approach to music emotion recognition,'' \emph{IEEE Trans. Audio, Speech, Lang. Process.}, vol.~16, no.~2, pp. 448--457, 2008.

\bibitem{emotion-rev}
------, ``Machine recognition of music emotion: {A} review,'' \emph{ACM Trans. Intelligent Systems and Technology}, vol.~3, no.~3, 2012.

\bibitem{mer}
J.~S.~G. Ca{\~{n}}{\'{o}}n, E.~Cano, T.~Eerola, P.~Herrera, X.~Hu, Y.-H. Yang, and E.~G{\'{o}}mez, ``Music emotion recognition: Toward new, robust standards in personalized and context-sensitive applications,'' \emph{IEEE Signal Process. Magzine}, vol.~38, no.~6, pp. 106--114, 2021.

\bibitem{bakker15}
D.~R. Bakker and F.~H. Martin, ``Musical chords and emotion: major and minor triads are processed for emotion,'' \emph{Cognitive, Affective, \& Behavioral Neuroscience}, 2015.

\bibitem{generationof}
Y.-C. Wu and H.~H. Chen, ``Generation of affective accompaniment in accordance with emotion flow,'' \emph{IEEE Trans. Audio, Speech, Lang. Process.}, 2016.

\bibitem{onperceived}
S.~Chowdhury and G.~Widmer, ``On perceived emotion in expressive piano performance: Further experimental evidence for the relevance of mid-level perceptual features,'' in \emph{Proc. ISMIR}, 2021.

\bibitem{audiofeatures}
R.~Panda, R.~Malheiro, and R.~P. Paiva, ``Audio features for music emotion recognition: {A} survey,'' \emph{IEEE Trans. Affective Computing}, 2020.

\bibitem{adata-driven}
A.~Aljanaki and M.~Soleymani, ``A data-driven approach to mid-level perceptual musical feature modeling,'' in \emph{Proc. ISMIR}, 2018.

\bibitem{TheEffects}
Y.~Hong, R.~K. Mo, and A.~Horner, ``The effects of mode, pitch, and dynamics on valence in piano scales and chord progressions,'' in \emph{Proc. ICMC}, 2018.

\bibitem{cp}
W.~Hsiao, J.~Liu, Y.~Yeh, and Y.-H. Yang, ``{Compound Word Transformer}: Learning to compose full-song music over dynamic directed hypergraphs,'' in \emph{Proc. AAAI}, 2021.

\bibitem{functionalharmony}
T.~Chen and L.~Su, ``Functional harmony recognition of symbolic music data with multi-task recurrent neural networks,'' in \emph{Proc. ISMIR}, 2018.

\bibitem{Controlling}
L.~N. Ferreira, L.~Mou, J.~Whitehead, and L.~H.~S. Lelis, ``Controlling perceived emotion in symbolic music generation with monte carlo tree search,'' in \emph{Proc. of AAAI (AIIDE Workshop)}, 2022.

\bibitem{Computer-Generated}
L.~N. Ferreira, L.~H.~S. Lelis, and J.~Whitehead, ``Computer-generated music for tabletop role-playing games,'' in \emph{Proc. of AAAI (AIIDE Workshop}, 2020.

\bibitem{AugmentedNet}
N.~N. L{\'{o}}pez, M.~Gotham, and I.~Fujinaga, ``Augmentednet: {A} roman numeral analysis network with synthetic training examples and additional tonal tasks,'' in \emph{Proc. ISMIR}, 2021.

\bibitem{notallroads}
G.~Micchi, M.~Gotham, and M.~Giraud, ``Not all roads lead to {Rome}: Pitch representation and model architecture for automatic harmonic analysis,'' \emph{TISMIR}, 2020.

\bibitem{ChordGNN}
E.~Karystinaios and G.~Widmer, ``Roman numeral analysis with graph neural networks: Onset-wise predictions from note-wise features,'' in \emph{Proc. ISMIR}, 2023.

\bibitem{Musictranscription}
B.~L. Sturm, J.~F. Santos, O.~Ben{-}Tal, and I.~Korshunova, ``Music transcription modelling and composition using deep learning,'' \emph{CoRR}, vol. abs/1604.08723, 2016.

\bibitem{mtharmonizer}
Y.~Yeh, W.~Hsiao, S.~Fukayama, T.~Kitahara, B.~Genchel, H.~Liu, H.~Dong, Y.~Chen, T.~Leong, and Y.-H. Yang, ``Automatic melody harmonization with triad chords: {A} comparative study,'' \emph{CoRR}, vol. abs/2001.02360, 2020.

\bibitem{class-octave}
Y.~Li, S.~Li, and G.~Fazekas, ``An comparative analysis of different pitch and metrical grid encoding methods in the task of sequential music generation,'' \emph{CoRR}, vol. abs/2301.13383, 2023.

\bibitem{CTRL}
N.~S. Keskar, B.~McCann, L.~R. Varshney, C.~Xiong, and R.~Socher, ``{CTRL:} {A} conditional transformer language model for controllable generation,'' \emph{CoRR}, vol. abs/1909.05858, 2019.

\bibitem{musicsimilarity}
M.~Mongeau and D.~Sankoff, ``Comparison of musical sequences,'' \emph{Computers and the Humanities}, vol.~24, no.~3, pp. 161--175, 1990.

\bibitem{c&e}
S.-L. Wu and Y.-H. Yang, ``{Compose \& Embellish}: Well-structured piano performance generation via a two-stage approach,'' in \emph{Proc. ICASSP}, 2023.

\bibitem{hooktheory}
``{HookTheory},'' \url{https://www.hooktheory.com/} [Accessed: (September 1, 2023)].

\bibitem{sheetsage}
C.~Donahue, J.~Thickstun, and P.~Liang, ``Melody transcription via generative pre-training,'' in \emph{Proc. ISMIR}, 2022.

\bibitem{skyline}
A.~L. Uitdenbogerd and J.~Zobel, ``Manipulation of music for melody matching,'' in \emph{Proc. ACM Multimed.}, 1998.

\bibitem{chorder}
J.~Chang, ``{Chorders},'' \url{https://github.com/joshuachang2311/chorder}.

\bibitem{KS-key}
C.~L. Krumhansl, ``Cognitive foundations of musical pitch,'' \emph{Oxford University Press}, 2001.

\bibitem{miditoolbox2016}
P.~Toiviainen and T.~Eerola, ``{MIDI} toolbox 1.1,'' https://github.com/miditoolbox/, 2016.

\bibitem{miditoolkit}
``{Midi\_Toolkit},'' \url{https://github.com/RetroCirce/Midi_Toolkit} [Accessed: (September 1, 2023)].

\bibitem{pop909}
Z.~Wang, K.~Chen, J.~Jiang, Y.~Zhang, M.~Xu, S.~Dai, and G.~Xia, ``{POP909:} {A} pop-song dataset for music arrangement generation,'' in \emph{Proc. ISMIR}, 2020.

\bibitem{transformer}
A.~Vaswani, N.~Shazeer, N.~Parmar, J.~Uszkoreit, L.~Jones, A.~N. Gomez, L.~Kaiser, and I.~Polosukhin, ``Attention is all you need,'' in \emph{Proc. NeurIPS}, 2017.

\bibitem{Transformer-XL}
Z.~Dai, Z.~Yang, Y.~Yang, J.~G. Carbonell, Q.~V. Le, and R.~Salakhutdinov, ``{Transformer-XL}: Attentive language models beyond a fixed-length context,'' in \emph{Proc. ACL}, 2019.

\bibitem{performer}
K.~M. Choromanski, V.~Likhosherstov, D.~Dohan, X.~Song, A.~Gane, T.~Sarl{\'{o}}s, P.~Hawkins, J.~Q. Davis, A.~Mohiuddin, L.~Kaiser, D.~B. Belanger, L.~J. Colwell, and A.~Weller, ``Rethinking attention with {Performers},'' in \emph{Proc. ICLR}, 2021.

\bibitem{sampling}
A.~Holtzman, J.~Buys, M.~Forbes, and Y.~Choi, ``The curious case of neural text degeneration,'' in \emph{Proc. ICLR}, 2019.

\bibitem{ATheoretical}
Z.~Fu, W.~Lam, A.~M. So, and B.~Shi, ``A theoretical analysis of the repetition problem in text generation,'' in \emph{Proc. AAAI}, 2021.

\bibitem{Emo-Harmonizer}
J.~Huang and Y.-H. Yang, ``Emotion-driven melody harmonization via melodic variation and functional representation,'' \emph{CoRR}, vol. abs/2407.20176, 2024.

\end{thebibliography}

%
%
%
%
%

\end{document}